\documentstyle[aps,twocolumn,prl]{revtex}

\begin{document}
\draft
\preprint{}
\title{
Spin resonance peak in Na$_{x}$CoO$_{2}\cdot y$H$_{2}$O
superconductors: A probe of the pairing symmetry }
\author{Jian-Xin Li$^{1,3}$ and Z. D. Wang$^{2}$}
\address{
1. National Laboratory of Solid States of Microstructure and Department of Physics, Nanjing University, Nanjing 210093, China\\
2. Department of Physics, The University of Hong Kong, Pokfulam Road, Hong Kong, China\\
3.The Interdisciplinary Center of Theoretical Studies, Chinese Academy of Sciences, Beijing 100080, China}
\maketitle

\begin{abstract}
Motivated by the recent discovery of superconductivity in
layered  Na$_{x}$CoO$_{2}\cdot y$H$_{2}$O compound, we investigate
theoretically the spin dynamics in the superconducting state with
 three possible $p_{x}+ip_{y}$-, $d+id'$-, and $f$-wave  pairing symmetries
 in two-dimensional triangular lattices. We find that a
spin resonance peak, which is elaborated to have a close relevance
to the relative phase of the gap function and the geometry of the
Fermi surface, appears in both in-plane and out-of-plane
components of the spin susceptibility $\chi$ for the spin-singlet
$d+id'$-wave pairing, while only in  the out-of-plane(in-plane)
component of $\chi$  for the spin-triplet $p_{x}+ip_{y}$-wave
($f$-wave) pairing. Because the pairing symmetry in this compound
is still hotly debated at present, this distinct feature may be
used to probe or determine it unambiguously  with future neutron
scattering experiments.

\end{abstract}

Recently, the superconductivity with $T_{c}\sim 5$K is found in
the CoO$_{2}$-layered material, Na$_{x}$CoO$_{2}\cdot
y$H$_{2}$O~\cite{tak}. This compound consists of
two-dimensional(2D) CoO$_{2}$ layers where Co atoms form a 2D
triangular lattice and is separated by a thick insulating layer
of Na$^{+}$ ions and H$_{2}$O molecules. Although its $T_{c}$ is
relatively low, it has still attracted much attention because it
shares some similarities with high-$T_{c}$ cuprates, in
particular it may be another unconventional superconductor
evolving from the doped Mott insulator. Therefore, the
investigation of this compound is expected to give insight on the
mechanism of unconventional superconductivity. So far, several
quite different theoretical
proposals~\cite{baska,kum,wang,ogata,tanaka,ikeda} for its
pairing symmetry, such as the spin-singlet $d+id'$-wave and the
spin-triplet $p_{x}+ip_{y}$-(or even $f$-)wave, have been put
forward. On the other hand, experimental results on the pairing
symmetry reported by different group have also been controversial,
even with the same experimental
method~\cite{koba,wad,fuji,ishida}.

In unconventional $d$- and $p$-wave superconductors,
 the gap function $\Delta ({\bf k})$ changes the sign(phase)
 around the Fermi surface and  thus would lead to zeros(nodes)
 in the superconducting(SC) energy gap $|\Delta ({\bf k})|$.
 Therefore, one can in principle determine the gap function and consequently
 the pairing symmetry, by measuring the distribution of the phase and/or
node positions. In practice, it is the node positions rather than
the phase that can be inferred in usual thermodynamic,
transport, and NMR experiments. Therefore, the probe of the node
position has been mostly used in the clarification of the pairing
symmetry in unconventional superconductors~\cite{li}. However, the
much debated pairing symmetries so far proposed for the SC state of
Na$_{x}$CoO$_{2}\cdot y$H$_{2}$O are the
broken-time-reversal-symmetry  $d+id'$- and $p_{x}+ip_{y}$- waves.
In this case, the energy gap $|\Delta ({\bf k})|=\sqrt
{\Delta^{2}_{d}({\bf k})+\Delta^{2}_{d'}({\bf k})}$ is nodeless.
So, a probe which is directly related to the phase is of special
importance for the determination of the pairing symmetry in this
compound as well as in other unconventional superconductors. 
This is evident by noting the crucial impact of the
tricrystal phase-sensitive experiment~\cite{tsuei} on the
determination of the dominating $d_{x^{2}-y^{2}}$ pairing
symmetry in high-$T_{c}$ superconductors.
In this Letter, by noting that a spin resonance peak appears in
different components of the dynamical spin susceptibility $\chi$ for all three possible
pairing symmetries in the 2D triangular lattice
with the nearest-neighbor(n.n.) pairing
interaction, we show that the
identification of the spin resonance peak in the SC state, which
can be carried out by neutron scattering experiments, may also
provide an unambiguous clue to probe/determine the pairing
symmetry in this compound.  We elaborate that the occurrence of
the spin resonance peak in a specific component of the dynamical
spin susceptibility exclusively corresponds to a {\it definite}
change of the phase of $\Delta ({\bf k})$.

To handle both the spin-singlet and spin-triplet
superconductivity in the same model 
as well as to capture the essential physics of 
electron correlation, we employ
a phenomenological $t-U-V$ model~\cite{zhu,han} on 
a 2D triangular lattice, in which an
effective n.n. pairing interaction ($V$) is responsible for
superconductivity and an on-site Hubbard $U$ for the electron
correlation. After choosing the mean-field parameter
$\Delta_{ij}^{\pm}=V(<c_{i\uparrow}c_{j\downarrow}>\pm
<c_{i\downarrow}c_{j\uparrow}>)/2$, we can write the effective
Hamiltonian as,
\begin{eqnarray}
H_{eff}&=&-\sum_{<ij>,\sigma}[tc^{\dag}_{i\sigma}c_{j\sigma}+h.c.]+U\sum_{i}n_{i\uparrow}n_{i\downarrow}  \nonumber \\
& &+\sum_{<ij>} [\Delta_{ij}^{\pm}(c^{\dag}_{i\uparrow}c^{\dag}_{j\downarrow}
\pm c^{\dag}_{i\downarrow}c^{\dag}_{j\uparrow})+h.c.] 
\end{eqnarray}
where the upper sign is for the spin-triplet pairing state and
the lower sign for the spin-singlet pairing state. Note that the
actual superconducting pairing mechanism is still unclear at
present and thus is not specified in this phenomenological model.

In the  triangular lattice, the dispersion of quasiparticles is
\begin{equation}
\epsilon_{k}=-2t[\cos k_{x}+2\cos{k_{x}\over 2}\cos{\sqrt{3}k_{y}\over 2}]-\mu
\end{equation}
 For 2D triangular lattices with the n.n. superconducting pairing
interaction, only $d_{x^{2}-y^{2}}\pm id_{xy}$-wave,
 $p_{x}\pm ip_{y}$-wave, and $f$-wave pairing states may
 exist~\cite{note1} :
 i) $\Delta^{d+id'}_{k}=\Delta_{0}\{[\cos(k_{x})-\cos(k_{x}/2)\cos({\sqrt3}k_{y}/2)]+i{\sqrt3}
  \sin(k_{x}/2)\sin({\sqrt3}k_{y}/2)\}$,
  ii) $\Delta^{p_{x}+ip_{y}}_{k}=\Delta_{0}\{[\sin(k_{x})+\sin(k_{x}/2)\cos({\sqrt 3}k_{y}/2]+i{\sqrt3} \cos(k_{x}/2)\sin({\sqrt3}k_{y}/2)\}$,
  and  iii) $\Delta^{f}_{k}=\Delta_{0} \{\sin(k_{x})-2\sin(k_{x}/2)\cos({\sqrt 3}k_{y}/2)\}$.
  Given the attractive interaction $V$ and with $t<0$~\cite{baska,wang,han},
  the mean-field calculation of Eq.(1) shows that
  nearly degenerate singlet $d+id'$- and triplet $f$-wave solutions  are favored at doping $n=0.4$, while
  the triplet $p_{x}+ip_{y}$-wave is stable at $n=1.35$, where $n$ is the average electron number per site~\cite{han,note2}.
    In order to compare the results for
different pairing symmetries with roughly the same SC
gap, we have chosen $V=0.75t$ for $n=0.4$ and $V=1.7t$ for
$n=1.35$, which gives
$\Delta_{0}=0.015t$ for $f$-wave, $\Delta_{0}=0.014t$ for
$d+id$-wave ($n=0.4$) and $\Delta_{0}=0.15t$ for $p+ip$-wave
($n=1.35$). The effective on-site Hubbard interaction is assumed to be
$U=3t$~\cite{note3}.

The bare spin susceptibility is given by,
\begin{eqnarray}
\chi^0_{ij}({\bf q},\omega ) &=&\frac 1{4N}\sum_k  [{C^{-}_{ij}(k,q)
(F^{+}_{k,q}-1) \over \omega -\Omega^{+}_{k,q}+i\Gamma} \nonumber \\ 
& &-{C^{-}_{ij}(k,q)
(F^{+}_{k,q}-1)\over \omega +\Omega^{+}_{k,q}+i\Gamma} 
+ {2C^{+}_{ij}(k,q)F^{-}_{k,q} \over {\omega +\Omega^{-}_{k,q}+i\Gamma}}]
\end{eqnarray}
where the coherence factors are,
\begin{equation}
C^{\pm}_{ij}(k,q)=[1\pm {{\epsilon _{k}\epsilon _{k+q}+{\rm Re}(\Delta _{k}\Delta^{*} _{k+q})}\over E_{k+q}E_{k}}]
\end{equation}
for the spin-singlet pairing and if $ij=zz$(the out-of-plane component of $\chi$) 
for the spin-triplet pairing, and
\begin{equation}
C^{\pm}_{ij}(k,q)=[1\pm {{\epsilon _{k}\epsilon _{k+q}-{\rm Re}(\Delta _{k}\Delta^{*} _{k+q})}\over E_{k+q}E_{k}}]
\end{equation}
if $ij=+-$(the in-plane component of $\chi$) for
the spin-triplet pairing. $F^{\pm}_{k,q}=f(E_{k+q})\pm f(E_{k})$
and $\Omega^{\pm}_{k,q}=E_k\pm E_{k+q}$, with $E_{k} = \sqrt
{\epsilon^2_{k} + |\Delta_{k}|^{2}}$ and $f(E_k)$ the Fermi
distribution function. Near $T=0$, only the first term in Eq.(3)
with the coherence factor $C^{-}$, involving the creation of
quasiparticle pairs, contributes to the spin susceptibility. An
essential difference in the coherence factors between the
spin-singlet pairing(or the component $\chi_{zz}$ for
spin-triplet pairing) and the component $\chi_{+-}$ for
spin-triplet pairing is a sign difference in front of
Re$[\Delta_{k}\Delta_{k+q}^{*}]$. This sign difference is the key
point for our later discussions.

The many-body correction to the spin susceptibility is included by the random phase approximation~\cite{note3}. In this way,
the renormalized spin susceptibility is given by,
\begin{equation}
\chi_{ij}({\bf q},\omega)={ \chi^{0}_{ij}({\bf q},\omega)[1-U
\chi^{0}_{ij}({\bf q},\omega)}]^{-1}.
\end{equation}

The momentum dependences of the dynamical spin susceptibility for $\omega=0.02t$ in the SC 
state ($T=0.0001t$) are presented in Fig.1. For the $p_{x}+ip_{y}$-wave pairing, 
a peak near ${\bf Q}_{1}=0.94\times(2\pi/3,2\pi/{\sqrt 3})$ can be seen. 
While, for the $f$-wave and $d+id'$-wave pairings, a peak near ${\bf Q}_{2}=(0,{\sqrt 3}\pi/2)$ 
appears. An obvious feature, seen from the figure, is that these peaks  depend only on the 
doping density and is irrespective of the symmetry of the pairing spin state and 
the components of Im$\chi$. This is due to the fact that these peaks arise from the nesting of
 Fermi surface which is determined only by the doping. The nesting wavevectors ${\bf Q}_{1}$ 
and ${\bf Q}_{2}$ which correspond to these two peaks are denoted as dashed lines with arrows in Fig.2. We note that a much sharper
peak occurs around ${\bf q}=(0,0)$ for the in-plane component Im$\chi_{+-}$ in the case of the spin-triplet pairing. This peak already presents in the normal state(not shown here) and reflects an enhanced ferromagnetic fluctuations which may arise from the substantial density of state at the Fermi level. However, it is suppressed highly for spin-singlet pairing
and for the out-of-plane component in the spin-triplet pairing as shown in Fig.1. 

In Fig.3(a), we present the frequency dependence of Im$\chi$ at
${\bf Q}_{1}$  for the spin-triplet $p_{x}+ip_{y}$-wave pairing.
It is seen that a spin resonance peak occurs near $\omega=0.05t$
for the out-of-plane component Im$\chi_{zz}$, but  is absent for
the in-plane component Im$\chi_{+-}$. However, in sharp contrast,
 the spin
resonance peak appears in the in-plane component of Im$\chi_{+-}$
rather than in the out-of-plane component for the spin-triplet
$f$-wave pairing(Fig.3(b)). Moreover, the spin resonance peak can 
be found in both components for $d+id'$ pairing, via the relation 
${\rm Im}\chi_{+-}=2{\rm Im}\chi_{zz}$ which holds for a spin-singlet state.
These distinctly different
features for all three possible pairing states are significantly
important because they may be used as an unambiguous clue to
probe/determine experimentally the pairing symmetry in this
compound.

To understand our observation clearly, we plot the bare spin susceptibility
Im$\chi_{0}$ in Fig.4.
It is clear that a peak is evident at the spin gap edge followed
by a step-like decrease just below the gap edge in the channels
in which there is a spin resonance peak. Using the
Kramers-Kroenig relation, we will obtain a logarithmic
singularity in its real part Re$\chi _{0}$. Thus, the RPA correction will
further magnify this effect and lead to a sharp peak near the gap
edge. This indicates that a peak just above the spin gap edge is
the source of the spin resonance.

 From the BCS theory,  the density of states(DOS) is divergent just
above the SC gap edge, which is expected to affect some physical
quantities. However, the effect is limited by the coherence
factor. The most evident effect of the coherence factor [Eqs.(4)
and (5)] is for energies $E_{k}$ and $E_{k+q}$ near the
quasiparticle gap edge $\Delta_{0}$, in which it is either $\sim
0$ or $\sim 1$, depending on the relative sign of $\Delta_{k}$ and
$\Delta_{k+q}$. Specifically, $C^{-}$ is negligible unless
$\Delta_{k}$ and $\Delta_{k+q}$ are of opposite signs for the
spin-singlet pairing~\cite{anderson} and for the out-of-plane
component in the spin-triplet pairing, or of the same sign for
the in-plane component in the spin-triplet pairing. With these
general considerations, let us now address why a peak appears in
one component of the spin susceptibility, but is absent in
the other. In Fig.2, we plot the phase($+$ sign denotes the phase
$0$, $-$ sign the phase $\pi$) and node position(dotted lines)
for various terms of the three possible pairing symmetries. The
Fermi surface for $n=1.35$, where the $p_{x}+ip_{y}$-wave is
favored, is a circle around $(0,0)$ point. For either the $p_{x}$ or
$p_{y}$ term, the two half circles separated by the line nodes
will have the opposite(different) sign(phase) of the gap function
$\Delta_k$. Therefore, for the wave vector ${\bf Q}_{1}=0.94\times(2\pi/3,
2\pi/\sqrt 3)$, $\Delta_{k}$ and $\Delta_{k+Q_{1}}$ have the
opposite sign. According to Eqs.(4) and (5), the coherence factor
$C^{-}$ is appreciable for $ij=zz$ and vanishes for $ij=+-$. As a
result, the DOS peak shows up in the out-of-plane component
Im$\chi_{zz}$ and does not in the in-plane component
Im$\chi_{+-}$, as shown in Fig.4(a). Also for the spin-triplet
pairing but with the $f$-wave,  $\Delta_{k}$ and
$\Delta_{k+Q_{2}}$ connected by ${\bf Q}_{2}=(0,{\sqrt 3}\pi/2)$
are of the same sign(phase)[Fig.2]. Therefore, the DOS peak exists in
Im$\chi_{+-}$, instead of in Im$\chi_{zz}$[Fig.4(b)]. The
most definite demonstration of this argument can be found in the
case of the $d+id'$ wave, where an appreciable coherence factor
$C^{-}$[Eq.(4)] requires that $\Delta_{k}$ and $\Delta_{k+q}$
have the opposite sign. However, from Fig.2 one can see that,
though the $\Delta_{k}$'s connected by wave vector ${\bf Q}_{1}$
satisfy the requirement for the $d'$ term, those for the $d$ term
do not. To see their effect, we have calculated the results for
$d$ and $d'$ terms, separately. As shown in the inset of
Fig.3(b), we find no peak for the $d$-wave term, but a sharp peak
for the $d'$-wave term. Remembering that the term
Re$[\Delta^{d+id'}_{k}\Delta^{d+id'
*}_{k+q}]=\Delta^{d}_{k}\Delta^{d}_{k+q}+\Delta^{d'}_{k}\Delta^{d'}_{k+q}$,
one will expect that it is the effect of $d'$ term dominates for
the $d+id'$ wave. The only difference between the cases of $d$- and
$d'$-wave pairings is their sign (phase) of the gap function. So,
it shows definitely that the spin resonance peak depends {\it
uniquely} on the relative phase of the gap functions connected by
the transition wave vector, when the Fermi surface is given.
Therefore,  the spin resonance peak appears in specific
components of the dynamical spin susceptibility  with distinctly
different ways for the proposed possible pairing states. We
suggest that neutron scattering experiments can probe the above
mentioned spin resonance peak, as done for
high-$T_{c}$ cuprates~\cite{res}, and thus to identify the pairing
symmetry of Na$_{x}$CoO$_{2}\cdot y$H$_{2}$O superconductors.

Before concluding the paper, let us use the above idea to address
the anisotropic suppression of the spin response at ${\bf
q}=(0,0)$ shown in Fig.1. At ${\bf q}\sim (0,0)$, the two gap
functions connected by ${\bf q}$ will surely have the same phase.
So, the coherence factor $C^{-}$ will be negligible for the
spin-singlet pairing and the out-of-plane component of the
spin-triplet pairing, but it is not for the in-plane component.
Thus, comparing with that in the normal state and in the latter case,
the spin response in the former case is strongly suppressed.

In conclusion, we have found that the spin resonance peak exists in 
distinctly different ways 
for three possible pairing symmetries proposed
for newly discovered CoO$_{2}$ layer materials Na$_{x}$CoO$_{2}\cdot y$H$_{2}$O,
and suggested to use it as an unambiguous clue to probe/determine the pairing
symmetry in  future inelastic neutron scattering experiments.
Moreover, we have elaborated that the spin resonance peak 
 has a close relevance
to the relative phase of the gap function and the geometry of the
Fermi surface.

After the completion of this work,
we noticed  a preprint~\cite{tao} by Li and Jiang, in which
 a spin resonance peak 
was  predicted only for the spin-singlet $d+id'$-wave pairing 
in the framework of the $t-J$ model on
triangular lattices and  by using the slave-boson approach. 

 We are grateful to Q. Han for useful discussions.  The work was supported
  by the National Nature Science Foundation of
China (10074025,10021001), the Ministry of Science and Technology
of China (NKBRSF-G19990646), and the RGC grant of Hong Kong.

\newpage
\section*{FIGURE CAPTIONS}

Fig.1  Momentum dependence of the spin susceptibility Im$\chi$
with $\omega=0.02t$ for (a) $p_{x}+ip_{y}$-wave at  $n=1.35$, and
(b) $f$- and $d+id'$- waves at $n=0.4$. The momentum is scanned
along the path shown in the inset of Fig.1(a).

Fig.2 Fermi surface (think lines) and phase ($\pm$) of the gap
functions for three possible pairing symmetries. The dotted lines
denote the node positions which separate the regions with
different phases, and the dashed lines with arrows represent the
transition wave vectors ${\bf Q}_{1}$ and ${\bf Q}_{2}$ as indicated in Fig.1.

Fig.3 Frequency dependence of the spin susceptibility Im$\chi$ in the SC state
($T=0.0001t$): (a) for the $p_{x}+ip_{y}$-wave at  $n=1.35$ and ${\bf
Q}_{1}$, (b) for the $f$- and $d+id'$-
waves at $n=0.4$ and ${\bf Q}_{2}$. The inset
of Fig.(b) shows the results for the pure $d$ and $d'$ waves,
respectively.

Fig.4 Frequency dependence of the bare spin susceptibility
$\chi_{0}$ in the SC state
($T=0.0001t$) calculated from Eq.(3):
(a)  for the $p_{x}+ip_{y}$-wave at  $n=1.35$ and ${\bf
Q}_{1}$, where the solid line indicates the
out-of-plane component $\chi^{0}_{zz}$ and the dashed line the
in-plane component $\chi^{0}_{+-}$;  (b)  for $f$- and $d+id'$-
waves at $n=0.4$ and ${\bf Q}_{2}$, where the
solid line indicates $\chi^{0}_{zz}$ of the $f$ wave, the dashed
line $\chi^{0}_{+-}$ of the $f$ wave and the dotted line
$\chi^{0}$ for $d+id'$ wave.


\begin{thebibliography}{99}
\bibitem{tak} K. Takada {\it et al.}, Nature {\bf 422}, 53 (2003).
\bibitem{baska} G. Baskaran, Phys. Rev. Lett. {\bf 91}, 097003 (2003).
\bibitem{kum} B. Kumar and B. S. Shastry, cond-mat/0304210.
\bibitem{wang} Q.H. Wang, D. H. Lee, and P. A. Lee, cond-mat/0304377.
\bibitem{ogata} M. Ogata, cond-mat/0304405.
\bibitem{tanaka} T. Tanaka and X. Hu, cond-mat/0304409.
\bibitem{ikeda} H. Ikeda, Y. Nisikawa, and K. Yamada, cond-mat/0308472.
\bibitem{koba} Y. Kobayashi, M. Yokoi, and M. Sato, cond-mat/0305649.
\bibitem{wad} T. Waki {\it et al.}, cond-mat/0306036.
\bibitem{fuji} T. Fujimoto {\it et al.}, cond-mat/0307127.
\bibitem{ishida} K. Ishida {\it et al.}, cond-mat/0308506.
\bibitem{li} For a review on unconventional superconductivity in Sr$_{2}$RuO$_{4}$, see
A. P. Mackenzie and Y. Maeno, Rev. Mod. Phys.{\bf 75}, 657 (2003) and also T. M. Rice, 
Physica C {\bf 341-348}, 41 (2000). A proposal for testing the node direction in layered organic superconductor
can be found in J. X. Li, Phys. Rev. Lett. {\bf 91}, 037002 (2003).
\bibitem{tsuei} C. C. Tsuei and J. R. Kirtley, Rev. Mod. Phys. {\bf 72}, 969 (2001).
\bibitem{zhu} J. X. Zhu and C. S. Ting, Phys. Rev. Lett. {\bf 87}, 147002 (2001).
\bibitem{han} Q. Han {\it et al.}, cond-mat/0306408.
\bibitem{note1}For a spin-triplet pairing state, a
three-component complex vector ${\bf d}({\bf k})=[d_{x}({\bf
k}),d_{y}({\bf k}),d_{z}({\bf k})]$ is adopted to represent
its spin dependence. In this way, the gap function can be written as,
$\Delta_{\alpha \beta}({\bf k})=[{\bf d}({\bf k}){\cdot} {\bf
\sigma}i\sigma_{2}]_{\alpha \beta}$, with ${\bf \sigma}$ are the
Pauli matrices. Following the argument in Ref.~\cite{tanaka}, we
assume that the pairing state is unitary and the ${\bf d}$ vector
is perpendicular to the cobalt plane, i.e., ${\bf d}||{\bf
z}||{\bf c}$. In this case, $\Delta_{k}=d_{z}({\bf
k})$. For a spin-singlet pairing, a single complex function
$\Delta_{k}$ is sufficient for the gap function.
\bibitem{note2}In the case of electron doping, the average electron number $n=1.35$($n=0.4$)
corresponds roughly to the experimental value in
Na$_{x}$CoO$_{2}\cdot y$H$_{2}$O with $x=0.35$($x=0.6$) for $t<0$($t>0$).
\bibitem{note3} We also tried other values in the range of $U=t \sim 3.5t$, where
the random phase approximation(RPA) is still valid. The conclusions obtained here do not change qualitatively. We note
that the weak coupling approache (RPA) can still give accurate results for the spin susceptibility even for the intermediate coupling if an effective(reduced) U is used, as shown by N. Bulut, D. J. Scalapino and S. R. White, Phys. Rev. B {\bf 47}, 2742 (1993).    
\bibitem{anderson} In the case of spin-singlet pairing such as the $d$ wave, this was previously pointed out by
Anderson {\it et al.} [Science{\bf 286}, 1154 (1995); Phys. Rev. Lett. {\bf 75}, 316 (1995)]
\bibitem{res} J. Rossat-Mignod {\it et al.}, Physica C {\bf 185}, 86 (1991); P. Dai {\it et al.}, Science {\bf 284}, 1344 (1999); P. Bourges {\it et al.}, Science {\bf 288}, 1234 (2000); H. He {\it et al.}, Science {\bf 295}, 1045 (2002) and references therein.
\bibitem{tao} T. Li and Y. J. Jiang, cond-mat/0309275.

\end{thebibliography}
\end{document}